\documentclass[12pt]{iopart}

\usepackage{citesort}

\expandafter\let\csname equation*\endcsname\relax
\expandafter\let\csname endequation*\endcsname\relax 
\expandafter\let\csname tr\endcsname\relax
\expandafter\let\csname mat\endcsname\relax

\usepackage{Symbols}
\usepackage{Shortcuts}
\usepackage{Text}

\usepackage{graphicx}
\usepackage{subfigure}
\usepackage{slashbox}

\newcommand{\gcv}[1]{\left\langle #1\right\rangle_{GCV}} 
\newcommand{\cbra}[1]{\bra{\tilde{#1}}} 
\newcommand{\cket}[1]{\ket{\tilde{#1}}} 

\def\gpe{g_{\perp}}
\def\gpa{g_{\parallel}}
\def\kT{k_BT}
\def\tfl{\tau_{FL}}
\def\Ham{\mathcal{H}} 
\def\Id{\mathds{1}} 
\def\Sm{\mathcal{S}} 

\def\Yb{\bar{\Y}} 
\def\Yd{\Y\dg} 

\begin{document}

\title[]{Crossover between strong and weak measurement in interacting many-body systems}

\author{Iliya Esin$^1$, Alessandro Romito$^2$, Ya. M. Blanter$^3$ and Yuval Gefen$^1$}
\address{$^1$Department of Condensed Matter Physics, The Weizmann Institute of Science, Rehovot 76100, Israel}
\address{$^2$Dahlem Center for Complex Quantum Systems and Fachbereich Physik, Freie Universit\"at Berlin, 14195 Berlin, Germany}
\address{$^3$Kavli Institute of Nanoscience Delft University of Technology Lorentzweg 1, 2628 CJ Delft, The Netherlands}

\vspace{10pt}

\begin{indented}
\item[]\today
\end{indented}

\begin{abstract}
Measurements with variable system-detector interaction strength, ranging from weak to strong, have been recently reported in a number of electronic nanosystems. In several such instances many-body effects play a significant role. Here we consider the weak-to-strong crossover for a setup consisting of an electronic Mach--Zehnder interferometer, where a second interferometer is employed as a detector. In the context of a conditional which-path protocol, we define a generalized conditional value (GCV), and determine its full crossover between the regimes of weak and strong (projective) measurement. We find that the GCV has an oscillatory dependence on the system-detector interaction strength. These oscillations are a genuine many-body effect, and can be experimentally observed through the voltage dependence of cross current correlations.
\end{abstract}

\vspace{2pc}
\submitto{\NJP}

\section{Introduction}
Measurement in quantum mechanics is inseparable from the dynamics of the system involved. The formal framework to describe quantum measurement, introduced by von Neumann \cite{Neumann1955}, allows to consider two limits: in the limit of strong system (S) –- detector (D)  coupling, the detector's final states are orthogonal. This is associated with the evasive notion of quantum collapse. In the other limit, that of weak (continuous) measurement of an observable (reflecting weak coupling between S and D \cite{Korotkov2001B}), the system is  disturbed  in a minimal way, and only partial information on the state of the latter is provided \cite{Clerk2010}. We note that this hindrance can be overcome, by resorting to a large number of repeated measurement (or a large ensemble of replica on which the same weak measurement is carried out).


Weak measurements, due to their vanishing back-action, can be exploited for quantum feedback schemes \cite{Zhang2005,Vijay2012} and conditional measurements. The latter is especially interesting for a two-step measurement protocol (whose outcome is called \ti{weak value} (WV) \cite{Aharonov1988}), which consists of a weak measurement (of the observable $\hat{A}$), followed by a strong one (of $\hat{B}$), $[\hat{A},\hat{B}]\ne0$. The outcome of the first is conditional on the result of the second (postselection). WVs have been observed in experiments \cite{Ritchie1991,Pryde2005,gorodetski2012,Groen2013,Tan2015,piccirillo2015}. Their unusual expectation values \cite{Aharonov1988,Romito2008,goggin2011,kofman2012} may be utilized for various purposes, including weak signal amplification \cite{Hosten2008,Dixon2009,Starling2009,Bruner2010,Starling2010,Zilberberg2011,Dressel2014,pang2015}, quantum state discrimination \cite{kocsis2011,lundeen2011,Zilberberg2013}, and non-collapsing observation of virtual states \cite{Romito2013}. The particular features of WVs rely on weak measurement, and are washed out in projective measurements. Understanding the relation and the crossover between these two tenets of quantum mechanics is therefore an important issue on the conceptual level.

The WV protocol perfectly highlights the difference between weak and strong (projective) measurements, thus providing a platform to study the crossover between the two. Indeed, within the two-step measurement protocol, it is possible to control the strength of the first measurement. This allows to define a \ti{generalized conditional value} (GCV), interpolating between WV and SV (\ti{strong value}). The latter, in similitude  to WV, refers to a 2-step measurement protocol. Unlike WV, in a SV protocol both steps consist of a strong measurement. The mathematical expression for GCV is depicted below in equation~\eqref{eq:GCVdefinition}. It amounts to the average of the first measurement's reading (whatever its strength is), conditional on the outcome of the second measurement. This has been studied in the context of single-degree-of-freedom systems \cite{Williams2008b,Lorenzo2012,Dressel2012A,Oehri2014}, where the WV-to-SV crossover is quite straightforward and is a smooth function of the interaction strength. We note that in experiments with electron nanostructures, interactions between electrons play a crucial role. A many-body theory of variable strength quantum measurement is called for. In many cases, the interaction strength can be controlled experimentally \cite{Groen2013,Weisz2013}.

In this letter, we demonstrate theoretically that interactions can modify this weak--to--strong crossover in a qualitative way, in particular, making it an oscillating function of the interaction strength. Conversely, these oscillations serve as a smoking gun manifestation of the many-body nature of the system at hand, and present guidelines for observing them as function of experimentally more accessible variables (e.g. the voltage bias). Our analysis sheds light on the relation between two seemingly very different descriptions of quantum measurement, with emphasis on the context of many-body physics.

Motivated by the two step WV protocol, we define the \ti{generalized conditional value} (GCV) of the operator $\hat{A}$ as an average shift of the detector, $\dl\hat{q}=\hat{q}-\av{\hat{q}}\at{g=0}$, during the measurement process, projected onto a postselected subspace by the projection operator, $\Pi_f$, and normalized by the bare S-D interaction strength, $g$. The GCV is given by
\begin{equation}
\gcv{\hat{A}}=\frac{\tr{\dl\hat{q}\hat{U}\dg\ro_0\hat{U}\Pi_{f}}}{g \tr{\hat{U}\dg\ro_0\hat{U}\Pi_{f}}},
\label{eq:GCVdefinition}
\end{equation}
where $\ro_0$ is the total density matrix which describes the initial state of S and D, and the time ordered operator $\hat{U}=\cT e^{-\frac{i}{\hb}\int_{-\infty}^{\infty}\cH^{SD}dt}$ describes the evolution in time of the whole setup during the measurement. Here, the system--detector coupling, $\cH^{SD}=-g w(t)\hat{p}\hat{A}$, with $w(t)$ -- the time window of the measurement; $\hat{q}$ and $\hat{p}$ are the ``position'' and ``momentum'' operators of the detector ($[\hat{q},\hat{p}]=i\hb$). We note that equation~\eqref{eq:GCVdefinition} provides the correct WV \cite{Aharonov1988} and SV \cite{Aharonov1964} in the respective limits ($g\ll1$, $g\gg1$). Our approach here is in full agreement with earlier analyses of quantum measurement in the context of single particle systems \cite{Williams2008b,Lorenzo2012,Dressel2012A,Oehri2014}.

\begin{figure}
  \centering
  \includegraphics[width=8.6cm]{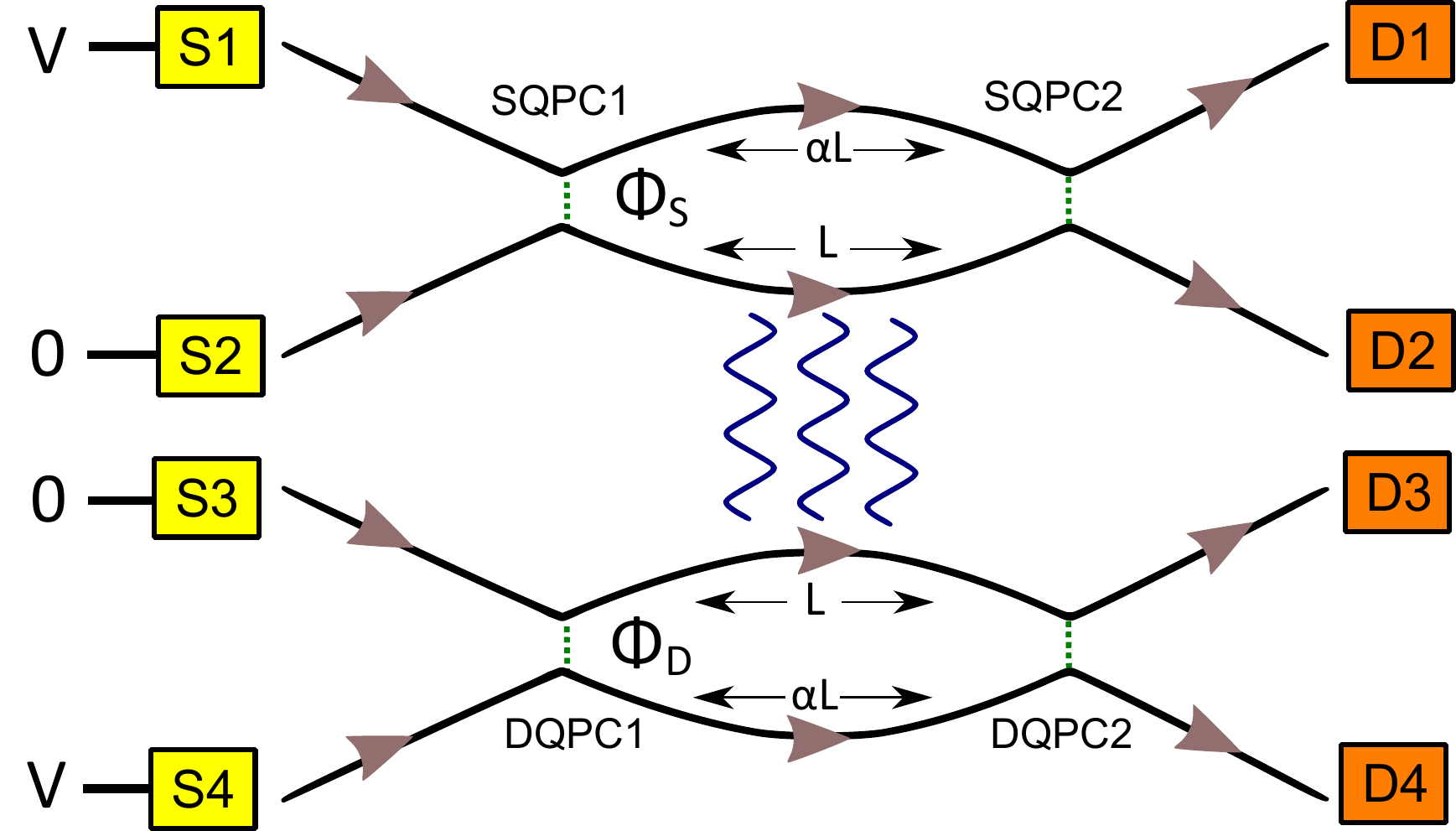}\\
  \caption{Two MZIs, the ``system'' and the ``detector'', coupled through an electrostatic interaction (wiggly lines). The sources S1 and S4 are biased by voltage V and the sources S2 and S3 are grounded. $\Phi_{S}$ and $\Phi_{D}$ are the magnetic fluxes through the respective MZIs. The lengths of the arms 1 and 2 between SQPC1 and SQPC2 are $\a L$ and $L$ respectively, and similarly for the detector's arms 3 and 4, as is shown in the figure. In the present analysis $\a=1$. \label{fig:TheSetup}}
\end{figure}

Our specific setup is depicted in figure~\ref{fig:TheSetup}. It consists of two Mach-Zehnder interferometers (MZIs), the ``system'' and the ``detector'' respectively, that are electrostatically coupled  \cite{Weisz2013,Chalker2007}. It is possible to tune the respective Aharonov-Bohm fluxes, $\Phi_{S}$  and $\Phi_{D}$ independently \cite{Weisz2013}.

\section{A two-particle analysis}
As a prelude to our analysis of a truly interacting many-body system, we briefly present an analysis of the same system on the level of a single particle in the system, interacting with a single particle in the detector. According to this (over)simplified picture, particles going simultaneously through the interacting arms 2 and 3 (cf. figure~\ref{fig:TheSetup}), gain an extra phase $e^{i\g}$ \cite{Dressel2012B,Neder2007a}, where $\g$ takes values in the range $[0,\p]$. First, we consider the intra-MZI operators, defined in a two-state single particle space, $\{\ket{m}\}$, with m=1,2 for the ``system'' (an electron propagating in arm 1 or 2) and similarly m=3,4 for the ``detector''. The dimensionless charge operator (measuring the charge between the corresponding quantum point contacts (QPCs)), in this basis has a form $Q_m=\ket{m}\bra{m}$. The transition through the p-th QPC is described by the scattering matrix
$\Sm_p=\mat{r_p & t_p\\-t^{\ast}_p & r_p}$, $p=1_s, 2_s, 1_d, 2_d$ \cite{Shpitalnik2008}. The entries $r_p$ and $t_p$ encompass information about the respective Aharonov-Bohm flux and for $p=2_s, 2_d$, about the orbital phase gained between the two QPCs. The dimensionless current operators at the source $(S1,S2)$ and the drain $(D1,D2)$ terminals of the system-MZI are given by $I_{Sm}=\Sm_{1_s}Q_m\Sm\dg_{1_s}$ and $I_{Dm}=\Sm\dg_{2_s}Q_m\Sm_{2_s}$ respectively, with $m=1,2$, and similarly for the detector with $m=3,4$ and employing the matrices $\Sm_{1_d}$ and $\Sm_{2_d}$.

In view of equation~\eqref{eq:GCVdefinition}, the initial state of the setup, which is described by the injection of two particles into terminals S1 and S4 respectively, can be written as the density matrix $\ro_0=I_{S1}\otimes I_{S4}$ operating in the two-particle product space, $\ket{m}\otimes\ket{n}$ ($m=1,2$, $n=3,4$). The corresponding dynamics is that of two particles propagating simultaneously through arms m and n. The interaction between the particles is described by the operator $\hat{U}=e^{i\g Q_2\otimes Q_3}$.
A positive reading of the projective measurement consists of the detection of a particle at D2, and is described by the projection operator $\Pi_f=I_{D2}\otimes \Id$. The detector reads the current at D3 ($\dl q$ of equation~\eqref{eq:GCVdefinition} corresponds to $\Id\otimes \dl I_{D3}$). Plugging these quantities into equation~\eqref{eq:GCVdefinition} yields an expression for the two-particle GCV (cf. \ref{sec:AppendixA}),
\begin{equation}
\gcv{Q_2}^{TP}=\frac{\av{I_{D2}\delta I_{D3}}}{\g\av{I_{D2}}}=\frac{1}{\g}\bR{\av{\dl I_{D3}}+\frac{\rav{I_{D2}I_{D3}}}{\av{ I_{D2}}}}.
\label{eq:GCVMZI}
\end{equation}
The averages are calculated with respect to the total density matrix after the measurement, $\langle \hat{O}\rangle= \tr{\hat{O}\hat{U}\dg\ro_0\hat{U}}$. We have defined $\dl I_{D3}\eqd I_{D3}-\av{I_{D3}}\at{\g=0}$, and $\rav{ I_{D2}I_{D3}}\eqd\av{I_{D2}I_{D3}}-\av{I_{D2}}\av{I_{D3}}$ is the irreducible current-current correlator. A straightforward calculation (cf. \ref{sec:AppendixB}) yields

\begin{equation}
\gcv{Q_2}^{TP}=\frac{4\sin\bR{\frac{\g}{2}}}{\g}\frac{\re{ie^{\frac{i\g}{2}}\av{I_{D2}Q_2}_0\av{\dl I_{D3}Q_3}}+\sin\bR{\frac{\g}{2}}\av{Q_2I_{D2}Q_2}_0\av{\dl Q_3I_{D3}Q_3}}
{\av{I_{D2}}_0+4\sin\bR{\frac{\g}{2}}\re{ie^{\frac{i\g}{2}}\av{I_{D2}Q_2}_0\av{Q_3}_0}+4\sin^2\bR{\frac{\g}{2}}\av{Q_2I_{D2}Q_2}_0\av{Q_3}_0}
\label{eq:GCVSingleParticle}
\end{equation}
where $\av{\hat{O}}_0\eqd\tr{\hat{O}\ro_0}$ is an average with respect to the non-interacting setup,
$\av{\dl I_{D3}Q_3}\eqd\av{I_{D3}Q_3}_0-\av{I_{D3}}_0\av{Q_3}_0$ and $\av{\dl Q_3I_{D3}Q_3}\eqd\av{Q_3I_{D3}Q_3}_0-\av{I_{D3}}_0\av{Q_3^2}_0$. This result shows a smooth and trivial crossover between the weak ($\g\to0$) and strong ($\g\to\p$) limits.  The specific form depends on the parameters of S and D (the magnitude of the inter-edge tunneling; the value of the Aharonov-Bohm flux). For some range of values (e.g., $t_{1_s}=t_{2_s}=t_{1_d}=t_{2_d}=0.1$, $\F_S/\F_0=0.99\p$, $\F_D=0$) the function is non-monotonic (but non-oscillatory), while for other values it is monotonic.


\section{A full many-body analysis}
The Hamiltonian $\cH=\cH^S+\cH^D+\cH^{SD}$ describes the system, the detector, and their interaction. The system's Hamiltonian consists of $\cH^{S}=\cH_0^{S}+\cH_T^{S}+\cH_{int}^{S}$, with
\begin{subequations}
\begin{eqnarray}
\cH_{0}^S=-iv_{F}\sum_{m=1}^{2}\int dx_m:\Y_{m}\dg(x_m)\dpa_{x_m}\Y_{m}(x_m):\\ \cH_{T}^S=\G_{1_s}\Y_{1}\dg(x_1^{1_s})\Y_{2}(x_2^{1_s})+\G_{2_s}\Y_{1}\dg(x_1^{2_s})\Y_{2}(x_2^{2_s})+h.c.\\
\cH_{int}^S=\sum_{m=1}^{2}\gpa\int dx_m:\bR{\Y_{m}\dg(x_m)\Y_{m}(x_m)}^2:.
\end{eqnarray}
\end{subequations}
Here $\G_{p}$ is the tunneling amplitude at QPC p and $x_m^p$ is the coordinate at QPC p on arm m. A similar expression holds for the ``detector'' MZI, $S\Leftrightarrow D$, with a summation over the chiral arms $m=3,4$. We next assume that the lengths of the interacting arms are equal, $x_2^{2_s}-x_2^{1_s}=x_3^{2_d}-x_3^{1_d}$. The S-D interaction Hamiltonian is
\begin{equation}
\cH^{SD}=\gpe\int dx_2\int dx_3\dl(x_2-x_3):\Y_{2}\dg(x_2)\Y_{2}(x_2)::\Y_{3}\dg(x_3)\Y_{3}(x_3):,
\end{equation}
where the normal ordering with respect to the equilibrium (no voltage bias) state is defined as  $:\Y\dg\Y:\eqd\Y\dg\Y-\braoket{0}{\Y\dg\Y}{0}$.

We are now at the position to construct the GCV for the actual many-body setup. We employ equation~\eqref{eq:GCVMZI} to define the many-body GCV of $Q_2$,
\begin{equation}
\gcv{Q_2}^{MB}=\frac{v_F}{\gpe}\bR{\av{\dl I_{D3}}+\frac{1}{\ta}\int_{-\ta/2}^{\ta/2}dt\frac{\rav{I_{D2}(t)I_{D3}(0)}}{\av{ I_{D2}}}},
\label{eq:GCVDiff}
\end{equation}
where the current operator is given by, $I(x,t)=ev_F:\Y\dg(x,t)\Y(x,t):$. We average over time $\ta\gg\frac{L}{v_F}$. The problem is now reduced to the calculation of average currents and a current-current correlator.  This is done perturbatively in the tunneling strength, but at arbitrary interaction parameter, employing the Keldysh formalism. In this limit expectation values are taken with respect to tunneling decoupled edge states. The current is,
\begin{equation}
\av{I_{D2}(x)}=-\frac{iev_{F}}{2}\sum_{p,q=\{1_s,2_s\}}\G_{p}\G_{q}\int \frac{d\w}{2\pi}G_{1,\a\be}(x-x_1^p,\w)
\hat{\g}_{\be\g}^{cl}G_{2,\g\dl}(x_2^p-x_2^q,\w)\hat{\g}_{\dl\e}^{cl}G_{1,\e\z}(x_1^q-x,\w)\hat{\g}_{\z\a}^{q},
\label{eq:CondCorrelator}
\end{equation}
 and the irreducible current-current correlator (cf. \ref{sec:AppendixC})

\begin{equation}
\begin{split}
&\frac{1}{\ta}\int_{-\ta/2}^{\ta/2}dt\rav{I_{D2}(x',t)I_{D3}(x,0)}=
\Bigg\{
\sum_{pqrs}\G_p\G_q\G_r\G_s
\frac{e^2 v_F^2}{2\ta}\int \frac{d\w_2d\w_3}{(2\p)^2} \times\\ &\times G_{1,\a\be}\bR{x'-x_1^p,\w_2-\frac{\bar{\w}}{2}}\hat{\g}_{\be\g}^{cl} G_{4,\h\q}\bR{x-x_4^r,\w_3+\frac{\bar{\w}}{2}}\hat{\g}_{\q\io}^{cl}
\tilde{M'}_{\io\ka\g\dl}\bR{x_2^p,x_2^q,x_3^r,x_3^s;\w_3,\w_2,\bar{\w}}\times \\
&\times\hat{\g}_{\ka\lm}^{cl} G_{4,\lm\m}\bR{x_4^{s}-x',\w_{3}-\frac{\bar{\w}}{2}}\hat{\g}_{\dl\e}^{cl} G_{1,\e\z}\bR{x_1^{q}-x,\w_{2}+\frac{\bar{\w}}{2}}\hat{\g}_{\z\a}^{q}\hat{\g}_{\m\h}^{q}
\Bigg\}
\at{\bar\w\to\frac{\sqrt{2\p}}{\ta}}+\Bigg\{...\Bigg\}\at{\bar\w\to-\frac{\sqrt{2\p}}{\ta}}.
\label{eq:CondCondCorrelator}
\end{split}
\end{equation}
Here $\{...\}$ reproduces the first part of the r.h.s, with $\bar\w\to\frac{\sqrt{2\p}}{\ta}$ replaced by $\bar\w\to-\frac{\sqrt{2\p}}{\ta}$, the summation is over $p,q=(1_s,2_s)$, $r,s=(1_d,2_d)$ and repeating indices; $\hat{\g}^{cl}=\mat{1 & 0\\0 & -1}$ and $\hat{\g}^{q}=\mat{1 & 0\\0 & 1}$ are the Keldysh $\hat{\g}$ matrices. $G_{m}$ is the fermionic propagator on the  m-th arm (cf. equation~\eqref{eq:Propagator}), and
\begin{equation}
\tilde{M}'(\w_2,\w_3)\eqd\tilde{M}(\w_2,\w_3)-G_2(\w_2)G_3(\w_3).
\end{equation}
Here $\tilde{M}_{\dl\g\be\a}(r_4,r_3,r_2,r_1)\eqd-\tav{\Y_{3,\dl}(r_4)\Yb_{3,\g}(r_3)\Y_{2,\be}(r_2)\Yb_{2,\a}(r_1)}$ is the collision matrix of two electrons in arms 2 and 3 (cf. \ref{sec:AppendixD}).

The expressions for the expectation values of equations~\eqref{eq:CondCorrelator} and \eqref{eq:CondCondCorrelator} can be represented diagrammatically in terms of the contributing processes. In these Feynman-Keldysh diagrams, each line corresponds to a propagator $G$ (cf. equation~\eqref{eq:Propagator}), and the vertices represent tunneling. The diagrams (to leading order in tunneling matrix elements) are depicted in figure~\ref{fig:Diagrams}. There are 16 diagrams contributing to the irreducible current-current correlator. The leading diagrams (figure~\ref{fig:Diagrams}~(b)) correspond to an electron in the system (going through arm 2) that maximally interacts with an electron in the detector (going through arm 3). \footnote[1]{For these diagrams the time of the two particles being inside the interaction region is maximal; the other diagrams are almost reducible (i.e., decoupled from each other), and are thus neglected.}

\begin{figure}
  \centering
  \includegraphics[width=8.6cm]{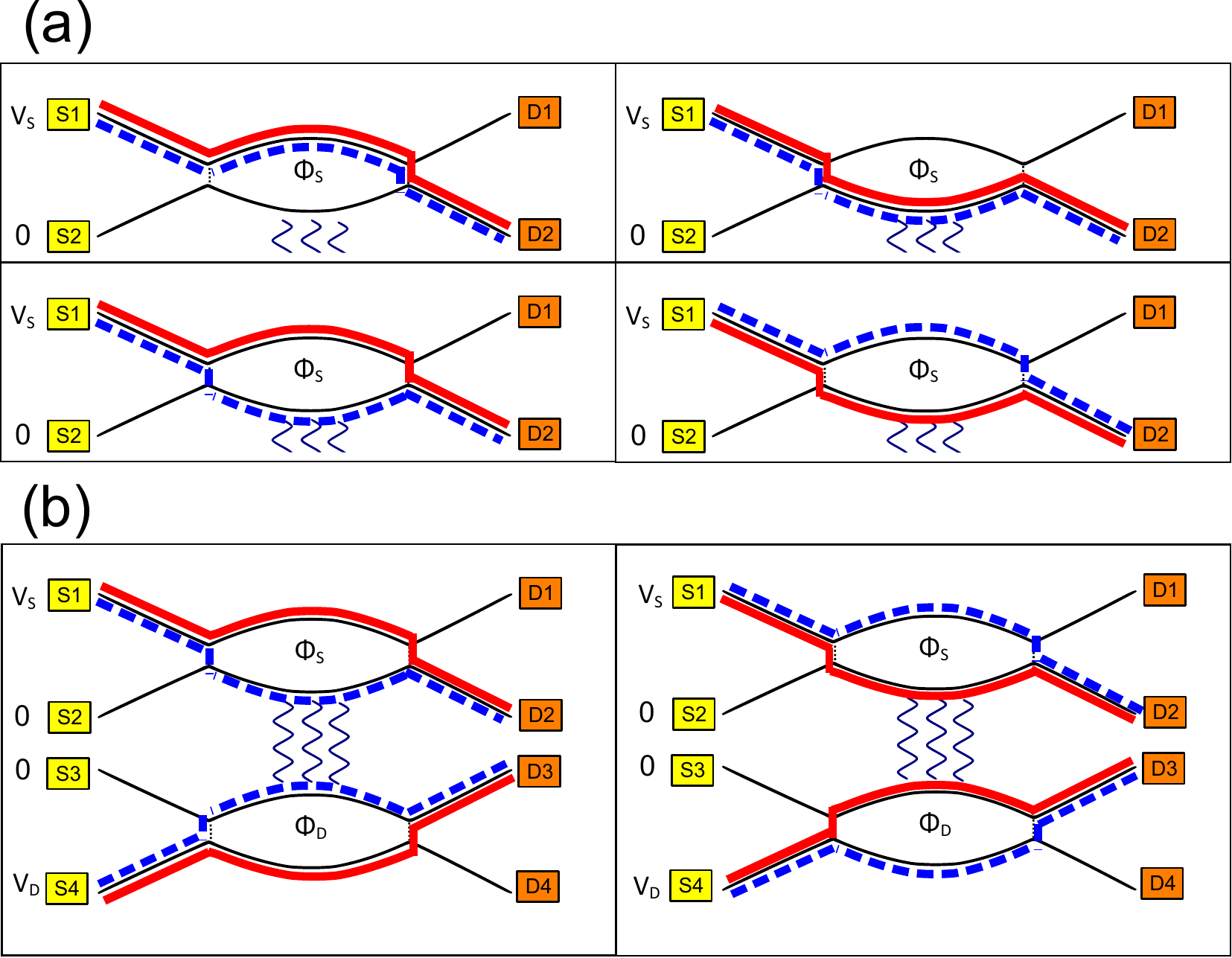}\\
  \caption{The relevant Feynman-Keldysh diagrams for the quantities in equations~\eqref{eq:CondCorrelator} and \eqref{eq:CondCondCorrelator} to leading order in tunneling matrix elements. ``Semi-classical'' paths of the particles are marked by solid lines (red) and dashed lines (blue), corresponding to forward and backward propagation in time (cf. equation~\eqref{eq:Propagator}). (a) The average current (equation~\eqref{eq:CondCorrelator}), $O(\G^2)$. Only the system part of the setup (cf. figure~\ref{fig:TheSetup}), while all degrees of freedom of the detector part have been integrated out. (b) The reducible current-current correlator (equation~\eqref{eq:CondCondCorrelator}), $O(\G^4)$. Only the 2 most contributing diagrams out of 16 are shown (4 were included in calculations).
  \label{fig:Diagrams}}
\end{figure}

Explicit evaluation of GCV requires the calculation of the single electron $G_m$ and the collision matrix $\tilde{M}$ \footnote[2]{As each channel is only slightly perturbed out of equilibrium, methods of equilibrium bosonization may be employed.}. We first compute the propagators on arms 2 ($G_2$) and 3 ($G_3$), where both the inter- and the intra-channel interaction is present. This yields
\begin{eqnarray}
G_{m,\be\a}(x,\w)=-\frac{i}{2v_{F}}\bS{F(\w)+\a\Q(x)-\be\Q(-x)}&&\times \nonumber \\*
\times e^{i\w\frac{x}{u}\x(\lm)}\int_{-1}^{1}\vs\bR{T\frac{|x|}{u}\lm,s}e^{is\w\frac{|x|}{u}\lm}ds&&,
\label{eq:Propagator}
\end{eqnarray}
where $\a,\be=\pm1$ are the Keldysh indices (in forward/backward basis), $x$ and $\w$ are the distance traveled by and the energy of the particle, and T is the temperature. We define the renormalized interaction $\lm=\bS{\frac{1}{u}\frac{2\gpe}{\p}-\bR{\frac{1}{u}\frac{2\gpe}{\p}}^{-1}}^{-1}$, $F(\w)=\tanh(\frac{\w}{2T})$, $\Q(x)$ is the Heaviside function, $\x(\lm)=\frac{2\lm^{2}}{\sqrt{4\lm^{2}+1}-1}$, and $\vs(A,s)=\frac{A}{\sqrt{\sinh\bS{\p A(1-s)}\sin\bS{\pi A(1+s)}}}$.

The propagators in channels 1 ($G_1$) and 4 ($G_4$) are obtained by substituting $\gpe=0$ in equation~\eqref{eq:Propagator}. This result recovers the simple non-interacting Green function with a renormalized velocity $u=v_{F}+\frac{2\gpa}{\p}$ due to intra-channel interaction.
The maximal interaction between channel 2 and 3 is at $\gpe=\frac{\pi}{2}u$ (instability point). Similarly to the two-particle analysis, here too the SV limit is reached at a finite value of the inter-channel interaction.

\section{Results}
\begin{figure}
  \centering
  \includegraphics[width=8.6cm]{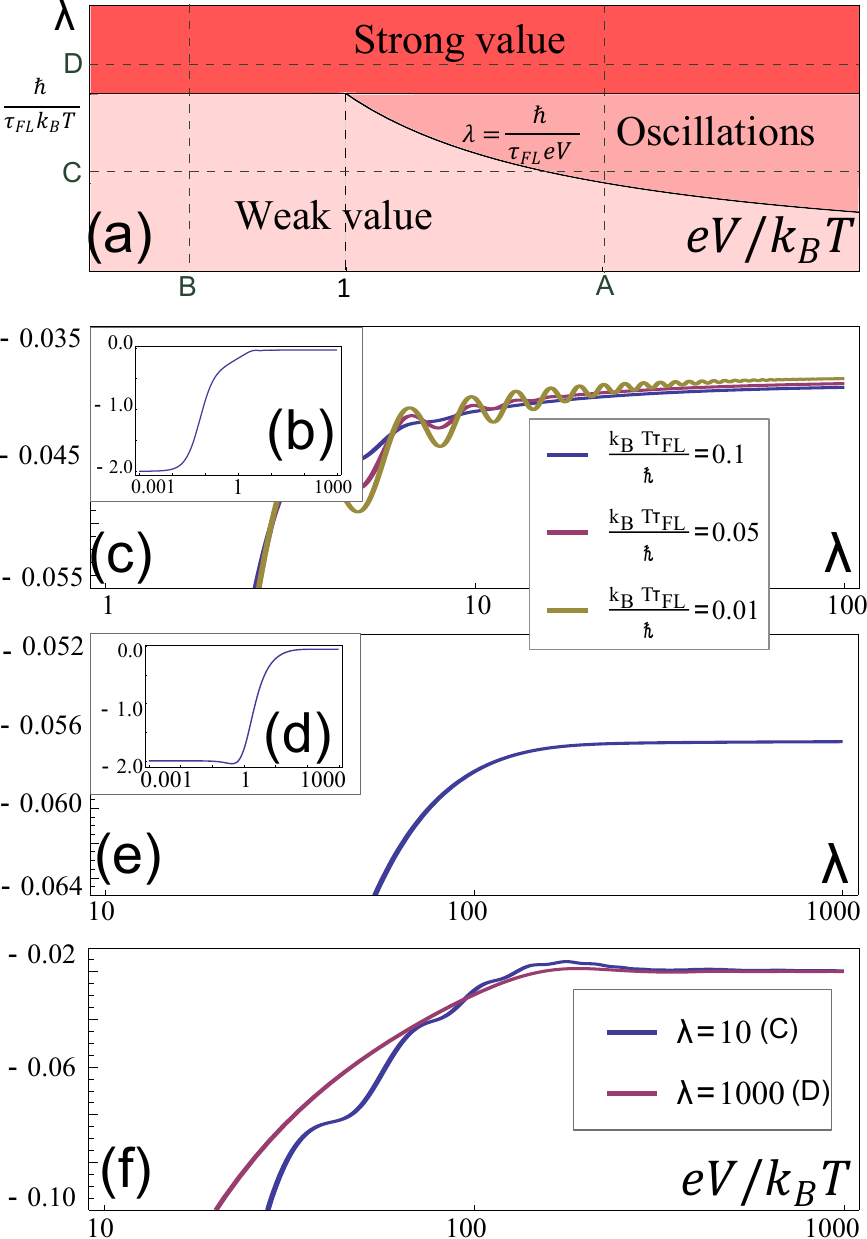}\\
  \caption{(a) The phase diagram in the low temperature regime, $\tfl\kT\ll\hb$. Regions with different qualitative behavior are depicted by different colors. The transition between weak and strong values in the high-voltage regime goes through an intermediate phase where the GCV displays oscillations as a function of the coupling constant. The latter feature is not present in the two-particle treatment of GCV (cf. equation~\eqref{eq:GCVSingleParticle}). (b) and (c). The normalized GCV, $\frac{\gcv{Q_2}^{MB}}{e^2 V/h}$, along the cuts A ($eV/\kT=100$), B ($eV/\kT = 0.001$) in (a). The zoom in (c) highlights the oscillatory behavior. (d) The oscillatory regime along A for various temperatures keeping $eV\tfl/\hb=1$. (e) The normalized GCV along the cuts C ($\lm$=10) and D ($\lm=1000$) of (a) with a zoom on the relevant oscillatory regime.  All the plots are for $\F_S/\F_0=0.99\p$, $\F_D=0$ at the low temperature phase, $\kT\tfl/\hb=0.01$ except of (d) where the temperatures are specified explicitly. \label{fig:GCVResult}}
\end{figure}

Plugging equations~\eqref{eq:Propagator} and equation~(S37) to equations~\eqref{eq:CondCorrelator} and \eqref{eq:CondCondCorrelator}, we obtain the final expression for the GCV in equation~\eqref{eq:GCVDiff}. The result is depicted in figure~\ref{fig:GCVResult}.
We identify a high temperature regime, $\tfl\kT\gg\hb$  ($\tfl$ is the time of flight through the interacting arm of MZI, $\tfl=\frac{L}{u}$), where the GCV is exponentially suppressed by the factor $e^{-\frac{\tfl\kT}{\hb}}$ due to averaging over an energy window $\eqs T$. In the opposite, low temperature limit, the phase diagram shows novel oscillatory behaviour. We plot the phase diagram of GCV in a parameter space spanned by the applied voltage normalized by the temperature ($eV/\kT$) and the renormalized interaction strength ($\lm$) (cf. figure~\ref{fig:GCVResult}). In the low voltage limit ($eV\ll\kT$) the size of the injected wave function is large compared with $L$. In this limit interaction effects should be less significant. The weak-to-strong crossover is smooth in similitude to the two particle result (cf. equation~\eqref{eq:GCVSingleParticle}). For $eV>\kT$, multiple particle interaction effects become important, and three different regimes are obtained as function of $\lambda$. Here, as function of increasing $\lm$, oscillatory behaviour ($\sim J_{0}\bR{\frac{\lm eV\tfl}{\hb}}$, where $J_{0}$ is the 0-th order Bessel function) of the crossover from WV to SV is predicted. The behavior of the GCV in the different regimes is summarized in a phase diagram in figure~\ref{fig:GCVResult}~(a), along with the dependence of the GCV on the interaction strength (figure~\ref{fig:GCVResult}~(b-d)) and voltage bias (figure~\ref{fig:GCVResult}~(e)).


\section{Discussion}
The oscillations found here and the physics of “visibility lobes” that was found experimentally \cite{Neder2006} and studied theoretically \cite{Sim2008,Neder2008,Kovrizhin2010} in the context of coherent transport through a MZI, are both related to interaction effects in an interferometry setup. To understand this similarity we employ a caricature semi-classical picture: a single particle wave-packet, whose energy components are in the interval $[0,eV]$, is  injected into the system MZI (arm 1 of figure~\ref{fig:TheSetup}). During its propagation through the interacting arm, its dynamics is affected by Coulomb interaction with the entire out-of-equilibrium Fermi sea of electrons inside the interaction region of the detector MZI (arm 3 of figure~\ref{fig:TheSetup}), producing a phase shift of the system’s wave-packet. When this single particle wave-packet interacts with a single electron in the detector (cf. the discussion preceding equation\eqref{eq:GCVMZI}), its phase shift is $0\le\g\le\p$. If the detector's arm consists of $N$ electrons, a phase shift of $N\g$ is produced, giving rise to oscillations as function of the interaction strength or $N$. More qualitatively: the number of background non-equilibrium electrons inside the detector MZI, $\av{N}=\frac{LeV}{2\p u}$ \cite{Sim2008,Neder2008,Kovrizhin2010}, splits into $n$ and $\av{N}-n$ in arms 3 and 4 respectively, with probability $P(n)=T^n R^{\av{N}-n} \binom{\av{N}}{n}$, $R=|r_{1_d}|^2$, $T=|t_{1_d}|^2$.

Neglecting, for the sake of this caricature, time dependent quantum fluctuations in the number of particles (we have treated those in full),  the incremental addition to the (system) wave packet action due to an electron in arm 2 interacting with $n$ background electrons in arm 3 is $\D S(n,t_0)=\frac{\gpe}{L}\int_{t_0}^{t_0+\tfl}n(t)dt$. Here $t_0\in [0,eV]$ is the injection time of an electron wave packet. The added phase to the single particle wave-function is: $\y\to\y e^{i\D S}$. It follows that the current at D2 per a specific $n$ is  $I_{D2}(n,t_0)=\frac{e^2V}{h}\bR{ R^2+T^2+2RT\cdot \re{e^{2\p i\frac{\F_S}{\F_0}}e^{i\D S(n,t_0)}}}$.
The mean current is a weighted average over all $\bC{n}$ and $t_0$, leading to a lobe structure. For example, when $\av{N}\ll1$, then $\D S(n)=\frac{\gpe\tfl n}{L}$, and the total current is $I=\frac{e^2V}{h}\bR{R^2+T^2+2RT\cdot D \re{e^{2\p i \frac{\F_S}{\F_0}+i\h_D}}}$, where $De^{i\h_D}=R+T e^{\frac{i\gpe\tfl\av{N}}{L}}$, which is periodic in $\gpe$ with a period of $\frac{(2\p)^2 u}{\tfl eV}$. We can repeat the same argument for the detector MZI and obtain the same lobe structure dependence there.

Measurements on setups consisting of two electrostatically coupled MZI have been reported \cite{Weisz2013}, albeit not in the context of the present work. By means of external gates one may control the magnitude of the coupling $\lm$. More accessible experimentally would  be to fix the distance between the MZIs and observe oscillations with $V$ at moderately low values of $\lm$.

The present analysis interpolates between two conceptually distinct views of measurement in quantum mechanics: the von Neumann projection postulate, and the continuous time evolution in the weak system-detector coupling limit. Admittedly these two views could be obtained as limiting cases of the same formalism. The analysis presented here demonstrates that the interpolation between the two is non-trivial. Oscillatory crossover is a unique feature of our many-body analysis. The setup chosen to demonstrate this SV-to-WV crossover consists of two coupled MZIs  (the ``system'' and the ``detector'').  Measurements on such a setup have been reported in the literature (see e.g., Ref.~\cite{Weisz2013}), with a considerable latitude of controlling the system-detector coupling. We conclude that our predictions are, then, within the realm of experimental verification.

\ack
We gratefully  acknowledge discussion with  Yakir Aharonov, Moty Heiblum, Itamar Sivan, Lev Vaidman and Emil Weisz. YG acknowledges the hospitality of the Dahlem Center for Complex Quantum Systems. This work is supported
by the GIF, ISF and DFG (Deutsche Forschungsgemeinschaft) grant RO 2247/8-1.

\appendix

\section{Derivation of the formula for two-particle GCV in terms of the irreducible correlation function \label{sec:AppendixA}}
Here we present an extended derivation of equation~\eqref{eq:GCVMZI}. The two-particle GCV of $Q_2$ is defined by,
\begin{equation}
\gcv{Q_2}^{TP}=\frac{\av{I_{D2}\dl I_{D3}}}{\g\av{I_{D2}}}=\frac{\av{I_{D2} \bR{I_{D3}-\av{I_{D3}}_0}}}{\g\av{I_{D2}}}
\label{eq:aGCVMZI1}
\end{equation}

This can be rewritten as,
\begin{equation}
\frac{\av{I_{D2}}\av{I_{D3}}-\av{I_{D2}}\av{I_{D3}}_0+\av{I_{D2}I_{D3}}-\av{I_{D2}}\av{I_{D3}}}{\g\av{I_{D2}}}
\label{eq:aGCVMZI2}
\end{equation}

which yields equation~\eqref{eq:GCVMZI},
\begin{equation}
\gcv{Q_2}^{TP}=\frac{1}{\g}\bR{\av{\delta I_{D3}}+\frac{\rav{I_{D2}I_{D3}}}{\av{I_{D2}}}}.
\label{eq:aGCVMZI3}
\end{equation}

\section{Strong--to--weak crossover of GCV for two particle system\label{sec:AppendixB}}
Here we present the derivation of GCV for two particle system (i.e. equation~\eqref{eq:GCVSingleParticle}). In accordance with equation~\eqref{eq:aGCVMZI1} we compute the current-current correlator $\av{I_{D2}I_{D3}}$ and the average current $\av{I_{D2}}$, defined with respect to the density matrix $\rho=e^{i\g Q_2 Q_3}I_{S1}\otimes I_{S4}e^{-i\g Q_2 Q_3}$,
\begin{equation}
\begin{split}
\av{I_{D2}I_{D3}}=&\tr{I_{D2}I_{D3}e^{i\g Q_2 Q_3}I_{S1}I_{S4}e^{-i\g Q_2 Q_3}}=\\
=&\tr{I_{D2}I_{D3}\bR{1+(e^{i\g}-1) Q_2 Q_3}I_{S1}I_{S4}\bR{1+(e^{-i\g}-1) Q_2 Q_3}}
\end{split}
\end{equation}
where in the last step we employed $e^{\g Q_2 Q_3}=1+\bR{e^{i\g}-1}Q_2 Q_3$ because the eigenvalues of $Q_i$ are only 0 or 1. Then,
\begin{equation}
\begin{split}
&\av{I_{D2}I_{D3}}=\\&=\av{I_{D2}}_0\av{I_{D3}}_0
\bR{1+\frac{4\sin\bR{\frac{\g}{2}}\re{ie^{\frac{i\g}{2}}\av{I_{D2}Q_2}_0\av{I_{D3}Q_3}_0}+4\sin^2\bR{\frac{\g}{2}}\av{Q_2I_{D2}Q_2}_0\av{Q_3I_{D3}Q_3}_0}{\av{I_{D2}}_0\av{I_{D3}}_0}
}
\label{eq:aAveID2ID3}
\end{split}
\end{equation}
where $\av{}_0$ denotes average with respect to the noninteracting setup ($\g\to0$). Similar calculation for $\av{I_{D2}}$ yields
\begin{equation}
\av{I_{D2}}=\av{I_{D2}}_0\bR{1+\av{Q_3}_0\frac{4\sin\bR{\frac{\g}{2}}\re{ie^{\frac{i\g}{2}}\av{I_{D2}Q_2}_0}+4\sin^2\bR{\frac{\g}{2}}\av{Q_2I_{D2}Q_2}_0}{\av{I_{D2}}_0}}.
\label{eq:aAveID2}
\end{equation}
Plugging equations~\eqref{eq:aAveID2ID3} and \eqref{eq:aAveID2} in equation~\eqref{eq:aGCVMZI1} yields an expression for a two particle GCV,
\begin{equation}
\gcv{Q_2}^{TP}=\frac{4\sin\bR{\frac{\g}{2}}}{\g}\frac{\re{ie^{\frac{i\g}{2}}\av{I_{D2}Q_2}_0\av{\dl I_{D3}Q_3}}+\sin\bR{\frac{\g}{2}}\av{Q_2I_{D2}Q_2}_0\av{\dl Q_3I_{D3}Q_3}}
{\av{I_{D2}}_0+4\sin\bR{\frac{\g}{2}}\re{ie^{\frac{i\g}{2}}\av{I_{D2}Q_2}_0\av{Q_3}_0}+4\sin^2\bR{\frac{\g}{2}}\av{Q_2I_{D2}Q_2}_0\av{Q_3}_0}.
\end{equation}
In the weak limit ($\g\to0$) this expression simplifies to
\begin{equation}
\lim_{\g\to0}\gcv{Q_2}^{TP}=2\re{\frac{i\av{I_{D2}Q_2}_0\av{\dl I_{D3}Q_3}}
{\av{I_{D2}}_0}},
\end{equation}
and in the strong limit ($\g\to\pi$),
\begin{equation}
\lim_{\g\to\infty}\gcv{Q_2}^{TP}=\frac{4}{\pi}\frac{\av{Q_2I_{D2}Q_2}_0\av{\dl Q_3I_{D3}Q_3}-\re{\av{I_{D2}Q_2}_0\av{\dl I_{D3}Q_3}}}
{\av{I_{D2}}_0-4\re{\av{I_{D2}Q_2}_0\av{Q_3}_0}+4\av{Q_2I_{D2}Q_2}_0\av{Q_3}_0}.
\end{equation}

\section{Perturbative calculation of expectation values\label{sec:AppendixC}}
In this section we derive the expression for expectation values of the current and the current-current correlator. Employing a path integral formalism, a general formula for the expectation value of an operator $\hat{O}[\Yd,\Y]$ is,
\begin{equation}
\av{\hat{O}[\Yd,\Y]}=\frac{\int\mathfrak{D}[\Yb,\Y]\hat{O}[\Yb,\Y]e^{iS[\Yb,\Y]}}{\int\mathfrak{D}[\Yb,\Y]e^{iS[\Yb,\Y]}},
\label{eq:aGeneralCorrelator}
\end{equation}
where $S=S_0+S_{int}+S_T$ is the full action over the Schwinger-Keldysh contour with
\begin{equation}
S_0[\Yb,\Y]=\sum_{m=1}^4\int drdr'\Yb_{m,\a}(r)\breve{G}^{-1}_{m,\a\be}(r-r')\Psi_{m,\be}(r'),
\end{equation}
\begin{equation}
S_{int}[\Yb,\Y]=\sum_{m,n=1}^4\int dr \rho_{m,\a}(r)g_{mn}\hat{\h}^{cl}_{\a\be}\rho_{n,\be}(r)
\end{equation}
and
\begin{equation}
S_T[\Yb,\Y]=\sum_{m,n=1}^4\int drdr' \Yb_{m,\a}(r)\G_{mn}(r,r')\hat{\g}^{cl}_{\a\be}\Y_{n,\be}(r').
\end{equation}
where $\a$,$\be$ are the Keldysh indices in forward/backward basis, m,n are the wire indices, r denotes the spacial 2-vector (r=(x,t)), $\rho_{m,\a}(r)=\Yb_{m,\a}(r)\Y_{m,\a}(r)$ is the density of the particles, $\hat{\h}^{cl}_{\a\be}$ is the Keldysh matrix (cf. Table~\ref{tab:KeldyshEta}),
\begin{equation}
g_{mn}=\mat{\gpa&0&0&0\\0&\gpa&\gpe&0\\0&\gpe&\gpa&0\\0&0&0&\gpa},
\label{eq:agMatrix}
\end{equation}

\begin{equation}
\G_{mn}(r,r')=\mat{
0&\G_s(x,x')&0&0\\
\G_s\as(x,x')&0&0&0\\
0&0&0&\G_d\as(x,x')\\
0&0&\G_d(x,x')&0
}\dl(t-t')
\label{eq:aGammaMatrix}
\end{equation}
and $\G_s(x,x')=\G_{1_s}\dl(x-x^{1_s}_1)\dl(x'-x^{1_s}_2)+\G_{2_s}\dl(x-x^{2_s}_1)\dl(x'-x^{2_s}_2)$ and $\G_d(x,x')=\G_{1_d}\dl(x-x^{1_d}_3)\dl(x'-x^{1_d}_4)+\G_{2_d}\dl(x-x^{2_d}_3)\dl(x'-x^{2_d}_4)$. $\breve{G}^{-1}_{m,\a\be}(k,\w)$ is the inverse of the fermionic Green function for particles whose dynamics is described by $\Ham^S_0+\Ham^D_0$, which in $(k,\w)$ representation is given by \cite{Kamenev2011}
\begin{equation}
\begin{split}
&\breve{G}_{m,\be\a}(k,\w)=\frac{1}{2}\bS{\frac{F(\w)+\a}{\w-v_Fk+i\e}-\frac{F(\w)-\be}{\w-v_Fk-i\e}}.
\label{eq:aFreePropagator}
\end{split}
\end{equation}
Here we assume the setup was in thermal equilibrium with a temperature T (described by the fermionic population function $F(\w)=\tanh\bR{\frac{\w}{2T}}$ at the time $t\to-\infty$, when the tunneling $\G$, and the interaction g were adiabatically turned on.
By assuming small tunneling the action can be expanded in power series to desired order in $\G$, then equation~\eqref{eq:aGeneralCorrelator} gets a form,
\begin{equation}
\av{\hat{O}[\Yd,\Y]}=\frac{\sum_n \frac{1}{n!}\av{\hat{O}[\Yb,\Y](iS_{T}[\Yb,\Y])^n}_{\W}}{\sum_n \frac{1}{n!}\av{(iS_{T}[\Yb,\Y])^n}_{\W}}
\label{eq:aGeneralCorrelatorPert}
\end{equation}
where $\av{}_{\W}$ denotes averaging with respect to the action $S_0+S_{int}$.

\begin{table}
\centering
\begin{tabular}{|c||c|c|}
  \hline
  \backslashbox{$\chi$}{$\a,\be$} & $(+/-)$ & (cl/q)\\
  \hline\hline
  $+$ & $\hat{\g}^+_{\a\be}=\mat{1&0\\0&0}$ & $\hat{\g}^+_{\a\be}=\frac{1}{2}\mat{1&1\\1&1}$ \\
  $-$ & $\hat{\g}^-_{\a\be}\mat{0&0\\0&-1}$ & $\hat{\g}^-_{\a\be}=\frac{1}{2}\mat{1&-1\\-1&1}$\\
  \hline
  cl & $\hat{\g}^{cl}_{\a\be}=\mat{1&0\\0&-1}$ & $\hat{\g}^{cl}_{\a\be}=\mat{1&0\\0&1}$ \\
  q & $\hat{\g}^{q}_{\a\be}=\mat{1&0\\0&1}$ & $\hat{\g}^{q}_{\a\be}=\mat{0&1\\1&0}$ \\
  \hline
\end{tabular}
\caption{A list of Keldysh $\hat{\g}^{\chi}_{\a\be}$ matrices (for fermions) in different bases of  bosonic ($\chi$) indices and fermionic indices ($\a,\be$). \label{tab:KeldyshGamma}}
\end{table}
\begin{table}
\centering
\begin{tabular}{|c||c|c|}
  \hline
  \backslashbox{$\chi$}{$\a,\be$} & $(+/-)$ & (cl/q)\\
  \hline\hline
  $+$ & $\hat{\h}^+_{\a\be}=\mat{1&0\\0&0}$ & $\hat{\h}^+_{\a\be}=\frac{1}{2}\mat{1&1\\1&1}$ \\
  $-$ & $\hat{\h}^-_{\a\be}\mat{0&0\\0&-1}$ & $\hat{\h}^-_{\a\be}=\frac{1}{2}\mat{-1&1\\1&-1}$ \\
  \hline
  cl & $\hat{\h}^{cl}_{\a\be}=\mat{1&0\\0&-1}$ & $\hat{\h}^{cl}_{\a\be}=\mat{0&1\\1&0}$ \\
  q & $\hat{\h}^{q}_{\a\be}=\mat{1&0\\0&1}$  & $\hat{\h}^{q}_{\a\be}=\mat{1&0\\0&1}$ \\
  \hline
\end{tabular}
\caption{A list of Keldysh $\hat{\h}^{\chi}_{\a\be}$ matrices (for bosons) in different bases of  bosonic $\chi,\a$ and $\be$ indices. \label{tab:KeldyshEta}}
\end{table}

The current in a chiral system with linear dispersion is linearly proportional to the density ($\av{I}=e v_F \av{\rho}$). The expectation value of the density is obtained by weakly perturbing the system by a quantum potential probe $V^q$, which should be taken to zero at the end to restore causality \cite{Kamenev2011}. Therefore, we obtain an expression for the current measured at $Dm$ ($m=1,2,3,4$) (cf. figure~\ref{fig:TheSetup}), $$\av{I_{Dm}(x,t)}=-\frac{iev_F}{2} \tr{\tilde{G}_m(x,t;x,t)\hat{\g}^q},$$ where $\tilde{G}_{m,\be\a}(x,t;x,t)=-i\tav{\Y_{m,\be}(x,t)\Yb_{m,\a}(x,t)}$ is the fermionic Green function of the system (averaged with respect to the full action, $S$) at point (x,t) of the m-th arm. The trace is over the Keldysh indices, where $\hat{\g}^q$ is the Keldysh matrix (cf. Table~\ref{tab:KeldyshGamma}). For the sake of simplicity we compute first $\av{I_{D1}(x,t)}$ by expanding it to second (leading) order in $\G$. We then employ the current conservation to find $\av{I_{D2}}$, $\av{I_{D2}(x,t)}=I_0-\av{I_{D1}(x,t)}$, where $I_0=\frac{e^2}{h}V$. To this order, particle tunnels twice. We employ  equation~\eqref{eq:aGeneralCorrelatorPert} to expand $\tilde{G}$ in $S_{\G}$. This yields
\begin{equation}
\begin{split}
\av{I_{D2}(x,t)}=&\frac{iev_F}{2}\int dt_1dt_2 \sum_{p,q=\{1_s,2_s\}}\\
&\Big\{\G_{p}\as\G_{q}
G_{1,\a\be}(x-x_1^p,t-t_1)\hat{\g}_{\be\g}^{cl}G_{2,\g\dl}(x_1^p-x_2^q,t_1-t_2)\hat{\g}_{\dl\e}^{cl}
G_{1,\e\z}(x_1-x,t_1-t)\hat{\g}_{\z\a}^q\Big\}.
\end{split}
\end{equation}
Here
\begin{equation}
G_m(x,t)_{\be\a}=-i\tav{\Y_{m,\be}(x,t)\Yb_{m,\a}(0,0)}
\label{eq:aGreenFunctionDef}
\end{equation}
is the fermionic Green function averaged with respect to the interacting action, $S_0+S_{int}$. We perform Fourier transform over the time variable to obtain,
\begin{equation}
\begin{split}
\av{I_{D2}(x,0)}=\frac{iev_F}{2}\int \frac{d\w}{2\pi} \sum_{p,q=\{1_s,2_s\}}\G_{p}\as\G_{q}
G_{1,\a\be}(x-x_1^p,\w)\hat{\g}_{\be\g}^{cl}G_{2,\g\dl}(x_1^p-x_2^q,\w)\hat{\g}_{\dl\e}^{cl}
G_{1,\e\z}(x_1-x,\w)\hat{\g}_{\z\a}^q.
\end{split}
\end{equation}

To find the current-current correlator, we generalize the last procedure, employing $\rav{I_{D2} I_{D3}}=\rav{I_{D1} I_{D4}}$, to obtain,

\begin{equation}
\begin{split}
&\frac{1}{\ta}\int_{-\ta/2}^{\ta/2}dt\rav{I_{D1}(x',t)I_{D4}(x,0)}=
-\frac{e^{2}v_{F}^{2}}{\ta}\int_{-\ta/2}^{\ta/2}dt\sum_{pqrs}\G_p\as\G_q\G_r\as\G_s\int dt_1dt_2dt_3dt_4 \times\\ &\times G_{1,\a\be}\bR{x'-x_1^{p},t'-t_1}\hat{\g}_{\be\g}^{cl} G_{4,\h\q}\bR{x-x_4^{r},0-t_4}\hat{\g}_{\q\io}^{cl} \tilde{M'}_{\io\ka\g\dl}\bR{x_3^{s},x_3^{r},x_2^{q},x_2^{p};t_3,t_4,t_2,t_1}\times\\
&\times\hat{\g}_{\ka\lm}^{cl} G_{4,\lm\mu}\bR{x_4^{s}-x',t_3-0}\hat{\g}_{\dl\e}^{cl} G_{1,\e\z}\bR{x_1^{q}-x,t_2-t}\hat{\g}_{\z\a}^{q}\hat{\g}_{\mu\h}^{q}
\end{split}
\end{equation}
where $\tilde{M}'(r_4,r_3,r_2,r_1)\eqd\tilde{M}(r_4,r_3,r_2,r_1)-G_2(r_2-r_1)G_3(r_4-r_3)$. And
\begin{equation}
\begin{split}
&\tilde{M}_{\dl\g\be\a}(r_4,r_3,r_2,r_1)\eqd-\tav{\Y_{3,\dl}(r_4)\Yb_{3,\g}(r_3)\Y_{2,\be}(r_2)\Yb_{2,\a}(r_1)}
\label{eq:aCollisionMatrixDef}
\end{split}
\end{equation}
is the collision matrix. We perform Fourier transform over the time differences, such that $\w_2$ corresponds to $t_2-t_1$, $\w_3$ to $t_4-t_3$ and $\bar{\w}$ to $\frac{1}{2}(t_3+t_4)-\frac{1}{2}(t_1+t_2)$. Finally, it yields

\begin{equation}
\begin{split}
&\frac{1}{\ta}\int_{-\ta/2}^{\ta/2}dt\rav{I_{D1}(x',t)I_{D4}(x,0)}=
-\frac{e^{2}v_{F}^{2}}{\ta}\int_{-\ta/2}^{\ta/2}dt\sum_{pqrs}\G_p\as\G_q\G_r\as\G_s
\int \frac{d\bar{\w}d\w_{2}d\w_{3}}{(2\pi)^{3}} e^{i\bar{\w}t}\times\\
&\times G_{1,\a\be}\bR{x'-x_1^{p},\w_{2}-\frac{\bar{\w}}{2}}\hat{\g}_{\be\g}^{cl} \times\label{eq:aCurCurCorrelator}
\times G_{4,\h\q}\bR{x-x_4^{r},\w_{3}+\frac{\bar{\w}}{2}}\hat{\g}_{\q\io}^{cl}
\tilde{M'}_{\io\ka\g\dl}\bR{x_3^{s},x_3^{r},x_2^{q},x_2^{p};\w_{3},\w_{2},\bar{\w}}\times\\
&\times \hat{\g}_{\ka\lm}^{cl} G_{4,\lm\mu}\bR{x_4^{s}-x',\w_{3}-\frac{\bar{\w}}{2}}\hat{\g}_{\dl\e}^{cl} G_{1,\e\z}\bR{x_1^{q}-x,\w_{2}+\frac{\bar{\w}}{2}}\hat{\g}_{\z\a}^{q}\hat{\g}_{\mu\h}^{q}.\nonumber
\end{split}
\end{equation}

In order to find a simpler expression for the time integral over $\ta$, we denote the current-current correlator by $F(t)$: $F(t)=\rav{I_{D2}(x',t)I_{D3}(x,0)}$, and its Fourier transform $F(\bar{\w})$. equation~\eqref{eq:aCurCurCorrelator} can be written in these terms as
\begin{equation}
\bar{F}\eqd\frac{1}{\ta}\int_{-\ta/2}^{\ta/2}dtF(t)=\frac{1}{\ta}\int_{-\ta/2}^{\ta/2}dt\int\frac{d\bar{\w}}{2\pi}e^{i\bar{\w}t}F(\bar{\w}).
\label{eq:aAverageF}
\end{equation}
It is easy to find an expression for $F(\bar{\w})$ by comparing equations~\eqref{eq:aCurCurCorrelator} and \eqref{eq:aAverageF}. First, we write
$\frac{1}{2\ta}\bS{F(\bar{\w})+F(-\bar{\w})}=\frac{1}{4\ta}\int_{-\infty}^{\infty}\bS{F(t)+F(-t)}(e^{i\bar{\w}t}+e^{-i\bar{\w}t})dt.$ From the other hand we approximate the average by, $$\bar{F}\approx\frac{1}{2\ta}\int_{-\infty}^{\infty}\bS{F(t)+F(-t)}e^{-\pi(t/\ta)^2}dt$$ where we have assumed that $F(t)$ grows much slower than $e^{\pi(t/\ta)^2}$, and the antisymmetric part of $F(t)$ is cancelled by the averaging. By comparing the exponentials in the two equations we obtain $\bar{\w}=\frac{\sqrt{2\pi}}{\ta}$. Then $\bar{F}=\frac{1}{2\ta}\bS{F(\frac{\sqrt{2\pi}}{\ta})+F(-\frac{\sqrt{2\pi}}{\ta})}$.

\section{Calculation of the fermionic correlators\label{sec:AppendixD}}
\begin{figure}
  \centering
  \includegraphics[width=8.6cm]{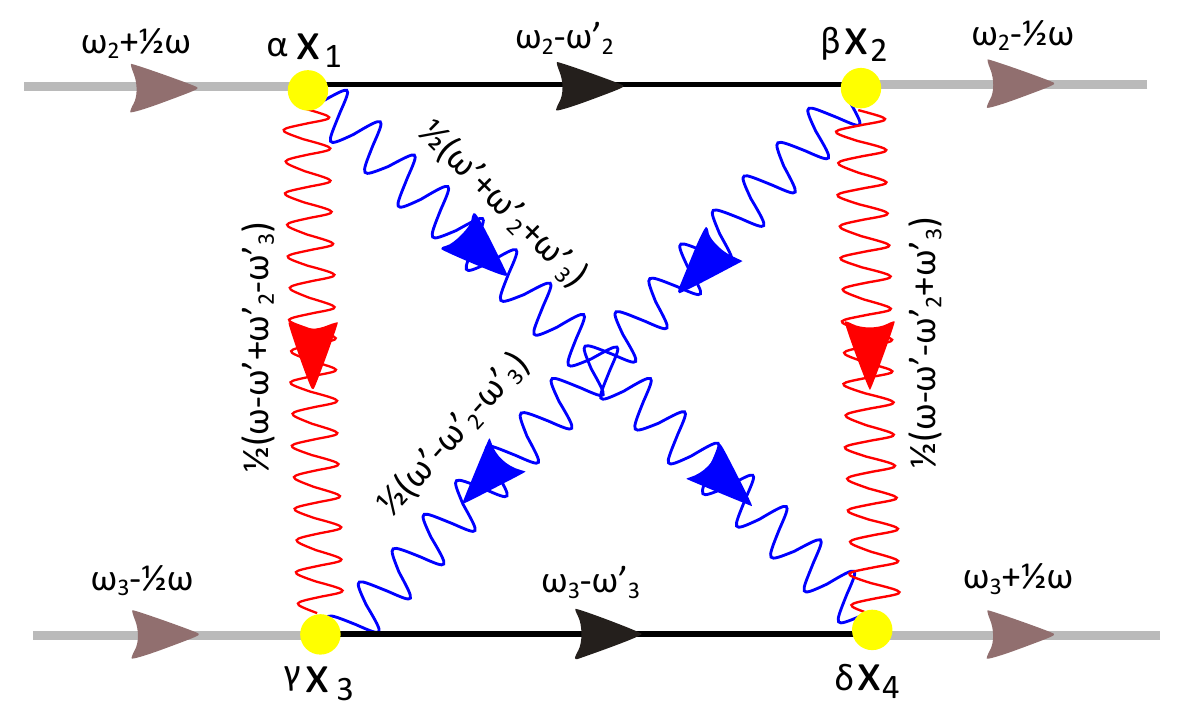}\\
  \caption{The collision matrix $\tilde{M}$ (cf. equation~\eqref{eq:aColMatrix}). A diagrammatic representation of the renormalized inelastic collision between two chiral fermions inside the interacting region. Straight lines correspond to fermionic Green functions (gray- outside the interacting region and black- inside). Wavy lines correspond to bosonic Green functions (red and blue for the two different types of bosons, cf. equation~\eqref{eq:aZparticleProp}). The vertices $x_1$,$x_3$ ($x_2$,$x_4$) correspond to the two entry (exit) points of the interaction region on the edges. The Keldysh indices ($\pm$) at these points are indicated by $\a,\be,\g,\dl$. Electrons enter the interacting region with energies $\w_{2}+\frac{1}{2}\w$ and $\w_{3}-\frac{1}{2}\w$  and exit with energies $\w_{2}-\frac{1}{2}\w$ and $\w_{3}+\frac{1}{2}\w$ respectively, exchanging energy $\w$ via 4 possible different bosons.}
  \label{fig:CollisionMatrix}
\end{figure}

Here we derive the expressions for the fermionic propagator (cf. equation~\eqref{eq:aGreenFunctionDef}) and the collision matrix (cf. equation~\eqref{eq:aCollisionMatrixDef}) averaged with respect to the action $S_0+S_{int}$, within an interacting arms (2,3) of MZI (the propagator in arms 1 and 4 can be found by taking $\gpe\to0$). In this calculation we employ the functional bosonization approach for system out of equilibrium \cite{Gutman2010,Mirlin2011}. We apply the Hubbard-Stratonovich transformation, and introduce the bosonic auxiliary field $\F$, writing an action $S_0+S_{int}$ as \cite{Lerner2002},
\begin{equation}
S_0+S_{int}[\Yb,\Y;\F]=\Yb G_{[\F]}^{-1}\Psi+\frac{1}{4}\F g^{-1}\F,
\label{eq:aFullAction1}
\end{equation}
with the notation,
$$\Yb G^{-1}_{[\F]}\Y=\sum_{m=2,3}\int drdr'\Yb_{m,\a}(r)G^{-1}_{[\F]m,\a\be}(r-r')\Y_{m,\be}(r')$$ where
$$G^{-1}_{[\F]m,\a\be}(r-r')=\breve{G}^{-1}_{m,\a\be}(r-r')-\hat{\g}^{\chi}_{\a\be}\F_{m,\chi}(r)\dl(r-r')$$ and

$$\F g^{-1}\F=\sum_{m,n=2,3}\int dr\F_{m,\a}(r) g_{mn}^{-1}\hat{\h}^{cl}_{\a\be}\F_{n,\be}(r).$$
where we implicitely sum over the Keldysh indices $\a,\be,\chi=\pm1$ (in forward/backward basis) and $g^{-1}_{mn}$ is the inverse of the $m,n=2,3$ submatrix of $g_{mn}$ (cf. equation~\eqref{eq:agMatrix}). Following the functional bosonization procedure \cite{Lerner2002}, we obtain a general expression for an n-fermion correlator,
\begin{eqnarray}
&\tav{\prod_i^n\Y_{a_i}(r_i)\Yb_{b_i}(q_i)}=\tav{\prod_i^n\Y_{a_i}(r_i)\Yb_{b_i}(q_i)}_0 e^{-\frac{1}{2}\tav{\bR{\sum_i^n\q_{a_i}(r_i)-\q_{b_i}(q_i)}^2}_{\F}},
\label{eq:aGeneralNCorrelator}
\end{eqnarray}
where $a,b=(\a,m)$ denote the Keldysh and the wire indices, $r,q=(x,t)$, $\av{}_0$ is the fermionic correlator with respect to the free action
\begin{equation}
S_0[\Yb,\Y]=\Yb\breve{G}^{-1}\Y,
\end{equation}
and $\av{}_{\F}$  is the $\F$-field correlator with respect to the action
\begin{equation}
S_{\F}[\F]=\frac{1}{4}\F g^{-1}\F+\F\hat{\Pi}\F
\label{eq:aPhiAction}
\end{equation}
respectively. Here $$\F\hat{\Pi}\F=\sum_{m=2,3}\int drdr'\F_{m,\a}(r) \hat{\Pi}_{m,\a\be}(r,r')\F_{m,\be}(r')$$ with the polarization matrix,
\begin{equation}
\hat{\Pi}_{m,\a\be}(r-r')=\frac{i}{2}\tr{\hat{\g}^{\a}\breve{G}_m(r-r')\hat{\g}^{\be} \breve{G}_m(r'-r)},
\end{equation}
where the trace is taken over the Keldysh fermionic indices \cite{Kamenev2011}. The $\q$ field is defined by
\begin{equation}
\q_{m,\a}(r)=-i\sum_{\be\g=\pm1}\int dr'G^B_{m,\a\be}(r-r')\hat{\h}^{cl}_{\be\g}\F_{m,\g}(r'),
\label{eq:aThetaFieldDefinition}
\end{equation}
where $G^B$ is the bosonic free Green function with linearized spectrum,
\begin{equation}
\begin{split}
&G^B_{m,\be\a}(k,\w)=\frac{1}{2}\bS{\frac{B(\w)+\a}{\w-v_Fk+i\e}-\frac{B(\w)-\be}{\w-v_Fk-i\e}}.
\end{split}
\end{equation}
The action for the $\F$ field (cf. equation~\eqref{eq:aPhiAction}) is quadratic due to Larkin-Dzyaloshinskii \cite{Larkin1974} theorem, therefore an exact expression for the $\F$-field correlator is
\begin{equation}
i\breve{Q}_{mn,\a\be}(r-r')\eqd\tav{\F_{m,\a}(r)\F_{n,\be}(r')}_{\F}=i\bR{g_{mn}^{-1}\hat{\h}^{cl}_{\a\be}\delta(r-r')+\delta_{mn}\hat{\Pi}_{m,\a\be}(r-r')}^{-1} \nonumber.
\end{equation}
We reduce the problem of finding an inverse of an infinite-dimensions matrix, inverting it to the finite (4) dimensions by Fourier-transforming it to a diagonal $(k,\w)$ basis.
Employing equation~\eqref{eq:aThetaFieldDefinition} we obtain the $\q$-field correlator,
\begin{equation}
\begin{split}
i\breve{K}_{mn,\a\be}(r-r')&\eqd\tav{\q_{m,\a}(r)\q_{n,\be}(r')}_{\F}=\\
&=-i\int dqdq' \bS{G^B(r-q)\hat{\h}^{cl}\breve{Q}(q-q')\hat{\h}^{cl}G^B(q'-r')}_{mn,\a\be},
\end{split}
\end{equation}

where we implicitly sum over the Keldysh and the wire indices. This yields,
\begin{equation}
\begin{split}
\breve{K}_{mn}=
\dl_{mn}\mat{B\bS{\breve{K}_{\parallel}^R-\breve{K}_{\parallel}^A}&\breve{K}_{\parallel}^R \\ \breve{K}_{\parallel}^A & 0}+\s^x_{mn}\mat{B\bS{\breve{K}_{\perp}^R-\breve{K}_{\perp}^A}&\breve{K}_{\perp}^R \\ \breve{K}_{\perp}^A & 0}
\end{split}
\end{equation}
where
\begin{equation}
\begin{split}
\breve{K}_{\parallel}^{R/A}(k,\w)=\frac{\pi}{k}\bS{\frac{1}{\w-v_{\rho}k\pm i\e}+\frac{1}{\w-v_{\s}k\pm i\e}-\frac{2}{\w-v_F k\pm i\e}},
\label{eq:aKcorrelatorPar}
\end{split}
\end{equation}
and
\begin{equation}
\breve{K}_{\perp}^{R/A}(k,\w)=\frac{\pi}{k}\bS{\frac{1}{\w-v_{\rho}k\pm i\e}-\frac{1}{\w-v_{\s}k\pm i\e}}.
\label{eq:aKcorrelatorPerp}
\end{equation}
Here, $v_{\rho}=u+\frac{2\gpe}{\pi}$, $v_{\s}=u-\frac{2\gpe}{\pi}$, with $u=v_{F}+\frac{2\gpa}{\pi}$.
We plug this result in equation~\eqref{eq:aGeneralNCorrelator} to compute the Green function (equation~\eqref{eq:aGreenFunctionDef}) and the collision matrix of the particles in arms 2 and 3 (equation~\eqref{eq:aCollisionMatrixDef}). The calculation requires transformation of equations~\eqref{eq:aKcorrelatorPar} and ~\eqref{eq:aKcorrelatorPerp} to real (x,t) space. Here we present the final result,
\begin{equation}
\begin{split}
G_{m,\be\a}(x,t)=-\frac{T}{2v_F}&\frac{1}{\sqrt{\sinh\bS{\pi T\bR{t-\frac{x}{v_{\rho}}+\frac{i}{\Lm}[\a\Q(t)-\be\Q(-t)]}}}} \times\\
\times&\frac{1}{\sqrt{\sinh\bS{\pi T\bR{t-\frac{x}{v_{\s}}+\frac{i}{\Lm}[\a\Q(t)-\be\Q(-t)]}}}}.
\label{eq:aPropagatorXT}
\end{split}
\end{equation}
Fourier-transforming the time coordinate yields,
\begin{equation}
\begin{split}G_{m,\be\a}(x,\w)=-\frac{i}{2v_{F}}\bS{F(\w)+\a\Q(x)-\be\Q(-x)} e^{i\w\frac{x}{u}\xi(\lm)}\int_{-1}^{1}\vs\bR{T\frac{|x|}{u}\lm,s}e^{is\w\frac{|x|}{u}\lm}ds
\label{eq:aPropagator}
\end{split}
\end{equation}
with the definitions $\lm=\bS{\frac{1}{u}\frac{2\gpe}{\pi}-\bR{\frac{1}{u}\frac{2\gpe}{\pi}}^{-1}}^{-1}$, $\xi(\lm)=\frac{2\lm^{2}}{\sqrt{4\lm^{2}+1}-1}$, and $\vs(A,s)=\frac{A}{\sqrt{\sinh\bS{\pi A(1-s)}\sinh\bS{\pi A(1+s)}}}$. For the sake of consistency check, $\lim_{\gpe\to0,\gpa\to0}G=\breve{G}$. And the collision matrix reads,
\begin{eqnarray}
\tilde{M}_{\dl\g\be\a}(x_4,x_3,x_2,x_1)=G_{3,\dl\g}(x_{43})G_{2,\be\a}(x_{21})\tilde{\z}^{(1)}_{\g\a}(x_{31})\tilde{\z}^{(1)}_{\dl\be}(x_{42})
\tilde{\z}^{(2)}_{\g\be}(x_{32})\tilde{\z}^{(2)}_{\dl\a}(x_{41})\nonumber
\end{eqnarray}
where,
$\tilde{\z}^{(1)}_{\be\a}(x,t)=\frac{\sqrt{\sinh\bS{\pi T\bR{t-\frac{x}{v_{\rho}}+\frac{i}{\Lm}[\a\Q(t)-\be\Q(-t)]}}}}{\sqrt{\sinh\bS{\pi T\bR{t-\frac{x}{v_{\s}}+\frac{i}{\Lm}[\a\Q(t)-\be\Q(-t)]}}}}$ and $\tilde{\z}^{(2)}_{\be\a}(x,t)=\bR{\tilde{\z}^{(1)}_{\be\a}(x,t)}^{-1}$. Fourier-transforming the time coordinates yields,
\begin{equation}
\begin{split}
&\tilde{M}_{\dl\g\be\a}(x_1,x_2,x_3,x_4,\w_{3},\w_{2},\w)=
\frac{1}{2}\int\frac{d\w'd\w'_{2}d\w'_{3}}{(2\pi)^{3}}G_{3,\dl\g}(x_{43},\w_{3}-\w'_{3})G_{2,\be\a}(x_{21},\w_{2}-\w'_{2})\times\\*
&\times\tilde{\z}^{(1)}_{\g\a}(x_{31},\frac{\w-\w'+\w'_{2}-\w'_{3}}{2})\tilde{\z}^{(1)}_{\dl\be}(x_{42},\frac{\w-\w'-\w'_{2}+\w'_{3}}{2})
\tilde{\z}^{(2)}_{\g\be}(x_{32},\frac{\w'-\w'_{2}-\w'_{3}}{2})\tilde{\z}^{(2)}_{\dl\a}(x_{41},\frac{\w'+\w'_{2}+\w'_{3}}{2})
\label{eq:aColMatrix}
\end{split}
,
\end{equation}
where we have used the short notation $x_{ij}=x_i-x_j$; $G$  is the single particle propagator given by equation~\eqref{eq:aPropagator}, and $\tilde{\z}^{(1/2)}_{\be\a}(x,\w)=2\pi\dl(\w)\cosh\bR{\frac{\pi Tx\lm}{u}}-2x\lm\tilde{Z}_{\be\a}^{(1/2)}(x,\w)$, where $\tilde{Z}^{(1/2)}_{\be\a}$ is given by,
\begin{equation}
\begin{split}
\tilde{Z}_{\be\a}^{(1/2)}(x,\w)=-\frac{i}{2u}[B(\w)+\a\Q(x)-\be\Q(-x)]e^{i\w\frac{x}{u}\xi(\lm)}\int_{-1}^{1}\kappa(T\frac{|x|}{u}\lm,s)e^{\pm is\w\frac{x}{u}\lm}ds,
\label{eq:aZparticleProp}
\end{split}
\end{equation}
where $B(\w)=\coth(\frac{\w}{2T})$ is the Bose function and $\kappa(A,s)=\sqrt{\frac{\sinh[\pi A(1+s)]}{\sinh[\pi A(1-s)]}}$.
equation~\eqref{eq:aColMatrix} has a pictorial interpretation, presented in figure~\ref{fig:CollisionMatrix}, according to which, the $\tilde{Z}$ particles are the dressed bosons that carry the interaction between the electrons.

\section{Passage of the electron through the MZI: a semiclassical picture\label{sec:AppendixE}}
\begin{figure}
  \centering
  \includegraphics[width=8.6cm]{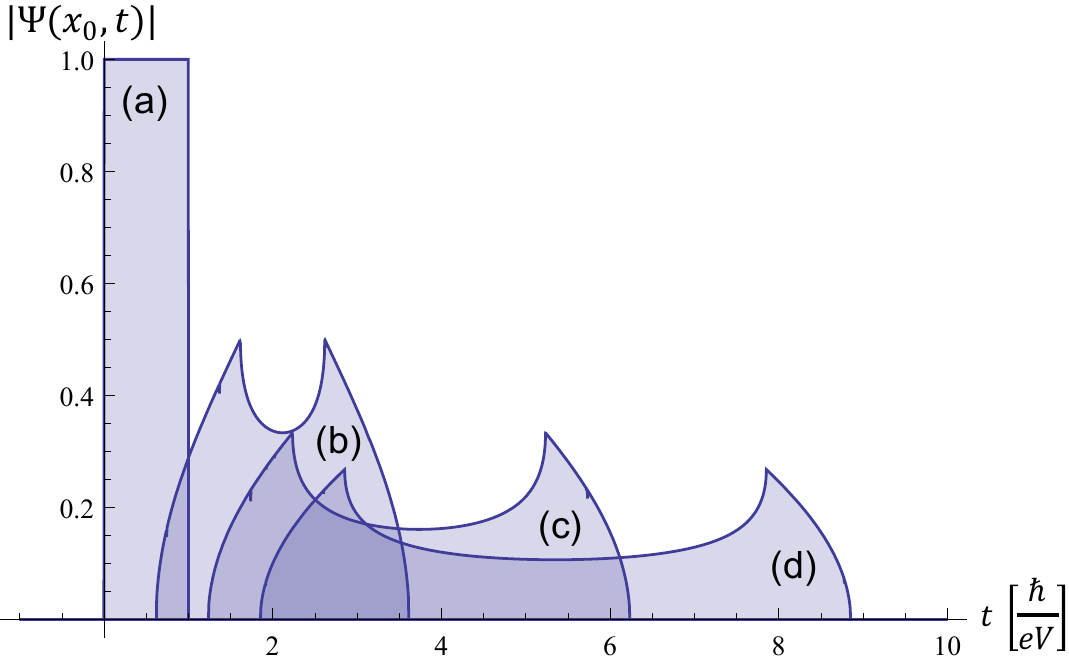}\\
  \caption{A propagation of the wave packet through an interacting arm of the MZI, at zero temperature, for $\lm=1$ for different points (a) $x_0=0$, (b) $x_0=\frac{u\hb}{eV}$, (c) $x_0=2\frac{u\hb}{eV}$, (d) $x_0=3\frac{u\hb}{eV}$. As can be derived from equation~\eqref{eq:aPropagatorXT}, the width of the wave packet is given by, $\D t=\frac{\hb}{eV}+\frac{2\lm x_0}{u}$.}
  \label{fig:Pulse}
\end{figure}

Here we present the propagation of a localized wave packet (according to a semiclassical picture) through an interacting arm of MZI, and derive the condition to be in the semiclassical regime.
We assume semiclassically a propagating rectangular shaped wave packet with a width $\sim\frac{\hb}{eV}$ in time domain (cf. figure~\ref{fig:Pulse}). The propagation of the single particle wave function can be derived by convolving the initial state with the retarded Green function,
\begin{equation}
\Y(x,t)=i\int G^R(x-x',t)*\Y(x',0) dx'.
\end{equation}
An expression for the zero temperature retarded Green function is (this is simply derived from equation~\eqref{eq:Propagator}).
\begin{equation}
G^R(x,t)=\frac{iu}{\pi\lm x v_F}\frac{\Pi\bR{\frac{u}{x\lm}(t-\frac{x\xi(\lm)}{u} )}}{\sqrt{1-\bR{\frac{u}{x\lm}(t-\frac{x\xi(\lm)}{u} )}^2}}
\label{eq:aPropagatorXT}
\end{equation}
where $\Pi(x)=\begin{cases}1 & -1<x<1\\0 & o.w.\end{cases}$ is a rectangle function. The wave packet at 4 different points is shown in figure~\ref{fig:Pulse}. We observe, the wave packet has been broadened as a result of the interaction, its width in time at different space points is given by $\D t(x_0)=\frac{\hb}{eV}+\frac{2\lm x_0}{u}$. The center of mass of the wave packet then propagates with velocity $v_{CM}=\frac{u}{\xi(\lm)}$. Consistent with the semiclassical picture, we require the width of the wave packet to be much smaller compared with the propagation time through the MZI, $\D t(L)\ll L/v_{CM}$. From this condition we deduce, $eV\gg\frac{\hb u}{L}$ and $\lm\ll 1$.

\section{General GCV for an N-state system\label{sec:AppendixF}}
Here we present a derivation of GCV for a general system with N-states being measured by a Gaussian detector. We show that the weak-to-strong crossover in such a case may be oscillatory with a bounded number of periods of the order of $O(N^2)$.
The initial state of the system is a mixed state, which is represented by the density matrix $\ro_s=\sum_{n,m}R_{nm}\ket{\a_n}\bra{\a_n}$. The detector is initialized in the zeroth coherent state (we denote the $\a's$ coherent state by $\cket{\a}$) such that its density matrix is $\ro_d=\cket{0}\cbra{0}$. We neglect the dynamics of the system and the detector assuming the measurement process was short in time compared to the typical timescales of the system and the detector. The coupling Hamiltonian is $\Ham_I=w(t)g \hat{A}(b^{\dagger}+b)$ with $b,b^{\dagger}$ are the ladder operators of the detector, $\hat{A}=\sum_n a_n\ket{\a_n}\bra{\a_n}$ and $w(t)$ is a window function around the time of the measurement. The post-selection is represented by the projection operator, $\Pi_f=\sum_{n,m}P_{nm}\ket{\a_n}\bra{\a_n}$. Plugging into equation~\eqref{eq:GCVdefinition} and considering, $\ro_{tot}=\ro_s\otimes\ro_d$ and $\delta q=b$, yields
\begin{equation}
\gcv{\hat{A}}=\frac{\sum_{n,m}a_n R_{nm}P_{mn}e^{-\frac{g^2}{2}(a_n-a_m)^2}}{\sum_{n,m} R_{nm}P_{mn}e^{-\frac{g^2}{2}(a_n-a_m)^2}}.
\end{equation}
The numerator and the denominator consist of sums of Gaussian (in g) functions, with different coefficients and prefactors. Each Gaussian is a monotonic function (for $g>0$), thus the maximal number of extremas in the weak-to-strong crossover ($g\in[0,\infty)$) is of the order of $O(N^2)$, where $N$ is the number of system's states.

\section{A full list of diagrams\label{sec:AppendixG}}
Figure~\ref{fig:Diagrams} depicts a full list of irreducible diagrams to fourth (leading) order in tunneling which should be taken in account for the current-current correlator. It is divided to diagrams with no flux dependence (cf. figure~\ref{fig:Diagrams}(a)), diagrams which are dependent on either $\F_S$ or $\F_D$ (cf. figures~\ref{fig:Diagrams}(b) and \ref{fig:Diagrams}(c)), and diagrams which are depend on both $\F_S$ and $\F_D$, cf. figure~\ref{fig:Diagrams}(d).

\begin{figure}
        \subfigure[one][Diagrams independent of the Aharonov-Bohm flux.]{\includegraphics[width=8.6cm]{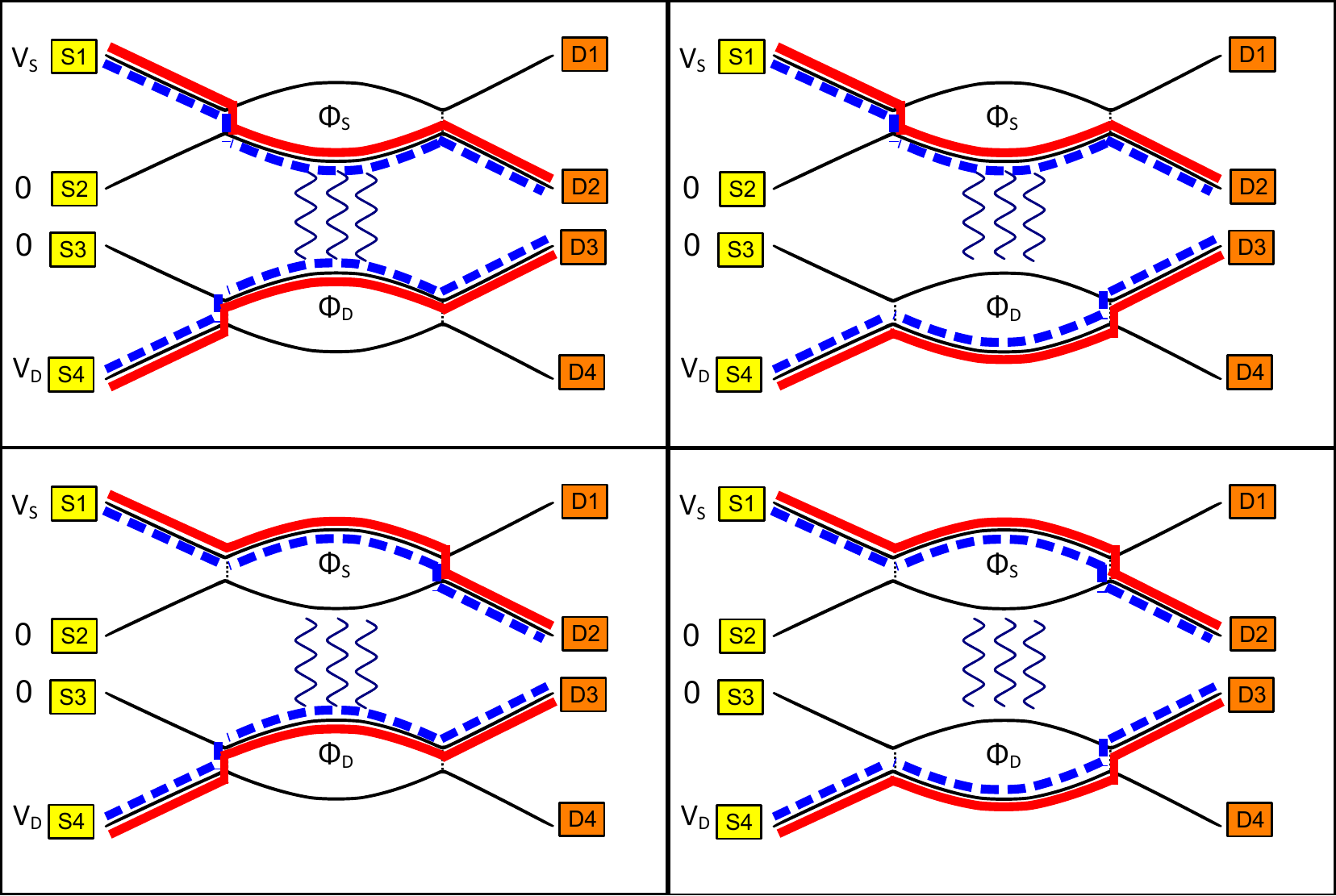}
                \label{subfig:Diagrams0}}
        \subfigure[two][Diagrams with $\Phi_S$ dependence.]{\includegraphics[width=8.6cm]{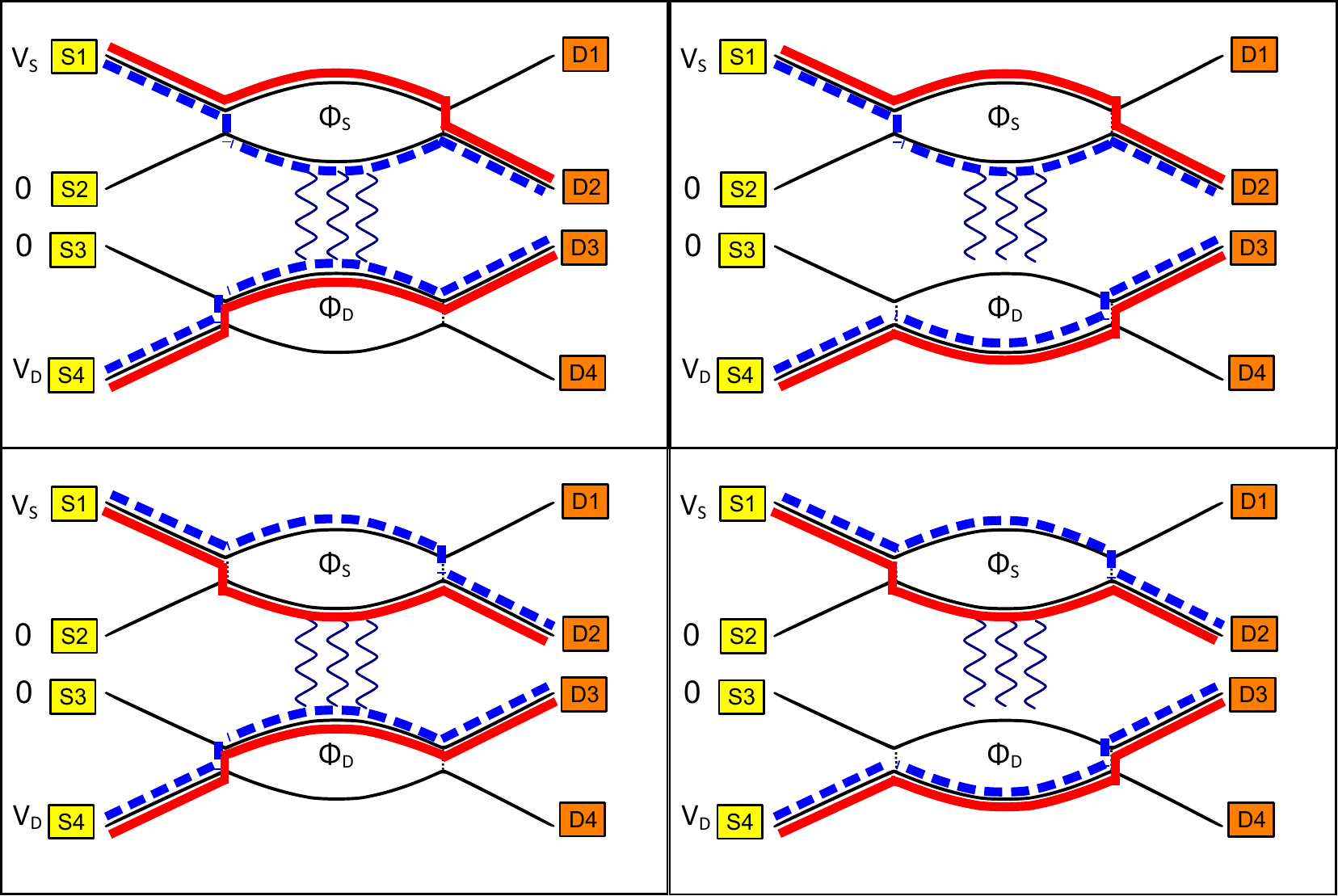}
                \label{subfig:DiagramsFS}}
        \subfigure[three][Diagrams with $\Phi_D$ dependence.]{\includegraphics[width=8.6cm]{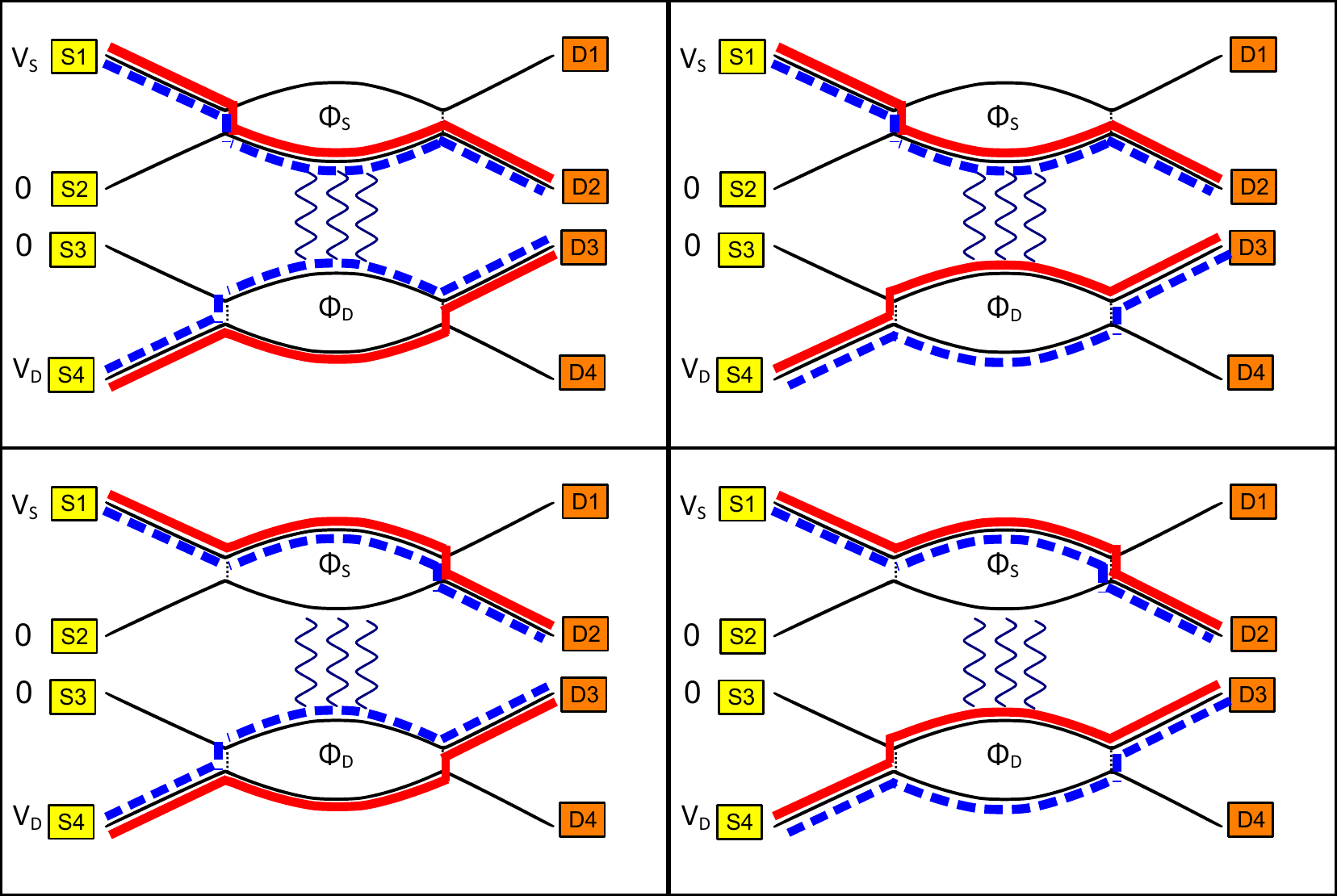}
                \label{subfig:DiagramsFD}}
        \subfigure[four][Diagrams depending on both $\Phi_S$ and $\Phi_D$.]{\includegraphics[width=8.6cm]{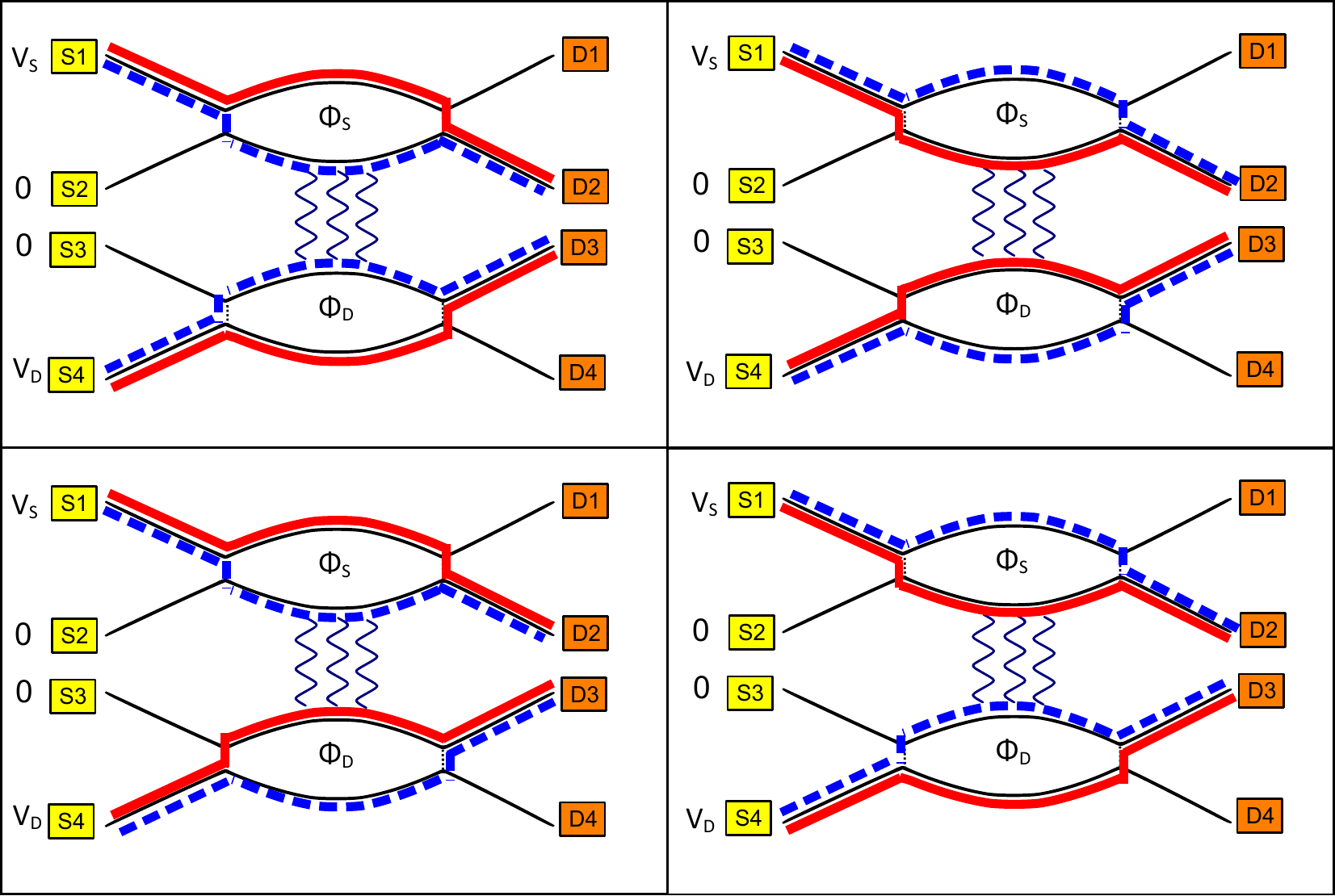}
                \label{subfig:DiagramsFDFS}}

        \caption{The full list of irreducible diagrams to fourth (leading) order in tunneling which should be taken in account for the current-current correlator (cf. equation~\eqref{eq:CondCondCorrelator}). “Semi-classical” paths of the particles are marked by solid lines (red) and dashed
lines (blue), corresponding to forward and backward propagation in time (cf. equations~(7) and (8)). The diagrams are divided to four groups by their Aharonov-Bohm flux dependence. The leading diagrams which were included in the calculation of the GCV, are in \ref{subfig:DiagramsFDFS}. \label{fig:Diagrams}}
\end{figure}

\end{document}